\documentclass[twocolumn]{article}
\usepackage[utf8]{inputenc} 
\usepackage[T1]{fontenc}    
\usepackage{hyperref}       
\usepackage{url}            
\usepackage{booktabs}       
\usepackage{amsfonts}       
\usepackage{nicefrac}       
\usepackage{microtype}      

\usepackage{graphicx}
\usepackage{subfig}

\usepackage{multirow}

\usepackage{array}
\usepackage{enumitem}

\usepackage{amssymb}
\usepackage{amsmath}

\usepackage{algorithm}      
\usepackage{algpseudocode}  
\usepackage{algorithmicx}   

\usepackage{xcolor}

\usepackage{authblk}

\usepackage{geometry}
\newcommand{\vx}{\mathbf{x}}
\newcommand{\vy}{\mathbf{y}}
\newcommand{\vd}{\mathbf{d}}
\newcommand{\va}{\mathbf{a}}
\newcommand{\vb}{\mathbf{b}}
\newcommand{\Sxx}{\Sigma_{xx}}
\newcommand{\Sxy}{\Sigma_{xy}}
\newcommand{\Syy}{\Sigma_{yy}}
\newcommand{\Shxx}{\hat{\Sigma}_{xx}}
\newcommand{\Shxy}{\hat{\Sigma}_{xy}}
\newcommand{\Shyy}{\hat{\Sigma}_{yy}}
\newcommand{\mA}{\mathbf{A}}
\newcommand{\mB}{\mathbf{B}}
\newcommand{\mAstar}{\mA^{\!*}}
\newcommand{\mBstar}{\mB^{*}}
\newcommand{\mU}{\mathbf{U}}
\newcommand{\mV}{\mathbf{V}}
\newcommand{\mT}{\mathbf{T}}
\newcommand{\mX}{\mathbf{X}}
\newcommand{\mY}{\mathbf{Y}}
\newcommand{\mXstar}{\mX^{\!*}}
\newcommand{\mYstar}{\mY^{\!*}}
\DeclareMathOperator*{\argmax}{arg\,max}
\DeclareMathOperator{\corr}{corr}
\DeclareMathOperator{\diag}{diag}
\DeclareMathOperator{\trace}{tr}
\DeclareMathOperator{\eigen}{eigen}
\DeclareMathOperator{\cholesky}{cholesky}
\DeclareMathOperator{\sgn}{sgn}
\newcommand{\overbar}[1]{\mkern 1.5mu\overline{\mkern-1.5mu#1\mkern-1.5mu}\mkern 1.5mu}
\newcommand{\ve}{\mathbf{e}}
\newcommand{\DCCA}{DCCA-2015}

\newcommand{\repo}{\url{https://github.com/CPJKU/cca_layer}}

\begin{document}

\title{End-to-End Cross-Modality Retrieval with \\ CCA Projections and Pairwise Ranking Loss
}



\author[1]{Matthias Dorfer}
\author[2]{Jan Schlüter}
\author[1]{Andreu Vall}
\author[1]{Filip Korzeniowski}
\author[1,2]{Gerhard Widmer}
\affil[1]{Department of Computational Perception, Johannes Kepler University Linz}
\affil[2]{The Austrian Research Institute for Artificial Intelligence}

\maketitle

\begin{abstract}
Cross-modality retrieval encompasses retrieval tasks where the fetched items
are of a different type than the search query, e.g., retrieving pictures relevant to a
given text query. The state-of-the-art approach to cross-modality retrieval relies on
learning a joint embedding space of the two modalities, where items from either
modality are retrieved using nearest-neighbor search. In this work, we
introduce a neural network layer based on Canonical Correlation Analysis (CCA)
that learns better embedding spaces by analytically computing projections that
maximize correlation. In contrast to previous approaches,
the CCA Layer (CCAL) allows us to combine existing objectives for embedding space learning, 
such as pairwise ranking losses, with the optimal projections of CCA. 
We show the effectiveness of our approach for cross-modality retrieval on three different scenarios (text-to-image, audio-sheet-music and zero-shot retrieval), surpassing both Deep CCA and a multi-view network using freely learned 
projections optimized by a pairwise ranking loss, especially when little training data
is available (the code for all three methods is released at: \repo).
%
\end{abstract}

\section{Introduction}
\label{sec:introduction}
Cross-modality retrieval is the task of retrieving relevant items of a
different modality than the search query (e.g., retrieving an image given a
text query). One approach to tackle this problem is to define transformations
which embed samples from different modalities into a common vector space.
We can then project a query into this embedding space, and retrieve, using
nearest neighbor search, a corresponding candidate projected from another
modality.

A particularly successful class of models uses parametric nonlinear
transformations (e.g., neural networks) for the embedding projections,
optimized via a retrieval-specific objective such as a pairwise ranking loss \cite{kiros2014unifying,socher2014grounded}.
This loss aims at decreasing the distance (a differentiable function such as
Euclidean or cosine distance) between matching items, while increasing it
between mismatching ones. Specialized extensions of this loss achieved
state-of-the-art results in various domains such as natural language
processing \cite{Hermann_2013_Multilingual}, image captioning 
\cite{Karpathy_2015_DeepVisual}, and text-to-image retrieval 
\cite{Vendrov_2016_OrderedEmbeddings}.

In a different approach, Yan and Mikolajczyk \cite{Yan2015DeepCorr} propose to learn a joint
embedding of text and images using Deep Canonical Correlation Analysis (DCCA) \cite{Andrew2013DCCA}.
Instead of a pairwise ranking loss, DCCA directly optimizes the correlation of learned latent representations of the two views.
Given the correlated embedding representations of the two views, it is possible to perform retrieval via cosine distance.
The promising performance of their approach is also in line with the findings of Costa et al. \cite{pereira2014role}
who state the following two hypotheses regarding the properties of efficient cross-modal retrieval spaces:
First, the embedding spaces should account for low-level cross-modal correlations and second, they should enable semantic abstraction.
In \cite{Yan2015DeepCorr}, both properties are met by a deep neural network --- learning abstract representations --- that is optimized with DCCA ensuring highly correlated latent representations.

In summary, the optimization of pairwise ranking losses yields embedding spaces that are useful for retrieval,
and allows incorporating domain knowledge into the loss function.
On the other hand, DCCA is designed to maximize correlation---which has already proven to be useful for cross-modality retrieval \cite{Yan2015DeepCorr}---but does not allow to use loss formulations specialized for the task at hand.

In this paper, we propose a method to combine both approaches in a way that retains their advantages.
We develop a \emph{Canonical Correlation Analysis Layer} (CCAL) that can be inserted into a dual-view neural network to produce a maximally correlated embedding space
for its latent representations.
We can then apply task specific loss functions, in particular the pairwise ranking loss, on the output of this layer.
To train a network using the CCA layer we describe how to backpropagate the gradient of this loss function to the dual-view neural network while relying on automatic differentiation tools such as \emph{Theano} \cite{Theano} or \emph{Tensorflow} \cite{abadi2016tensorflow}.
In our experiments, we show that our proposed method performs better than DCCA and models using
pairwise ranking loss alone, especially when little training data is available.
\begin{figure*}[ht!]
\centering
\subfloat[DCCA network maximizes correlation via Trace Norm Objective (\emph{TNO}).]{\label{fig:dcca_sketch}{\includegraphics[height=0.27\textwidth]{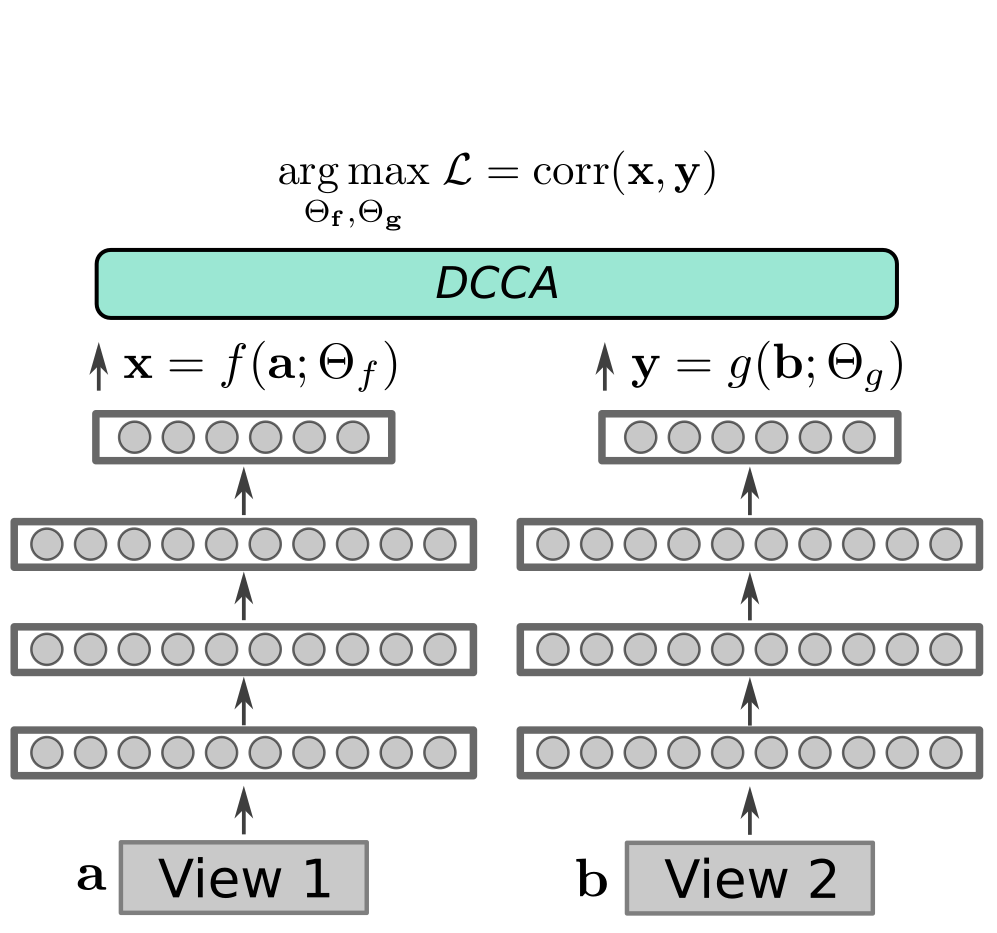} }}%
\quad
\subfloat[Freely-learned embedding projections optimized with ranking loss (\emph{Learned-$\mathcal{L}_{rank}$}).]{\label{fig:learned_sketch}{\includegraphics[height=0.27\textwidth]{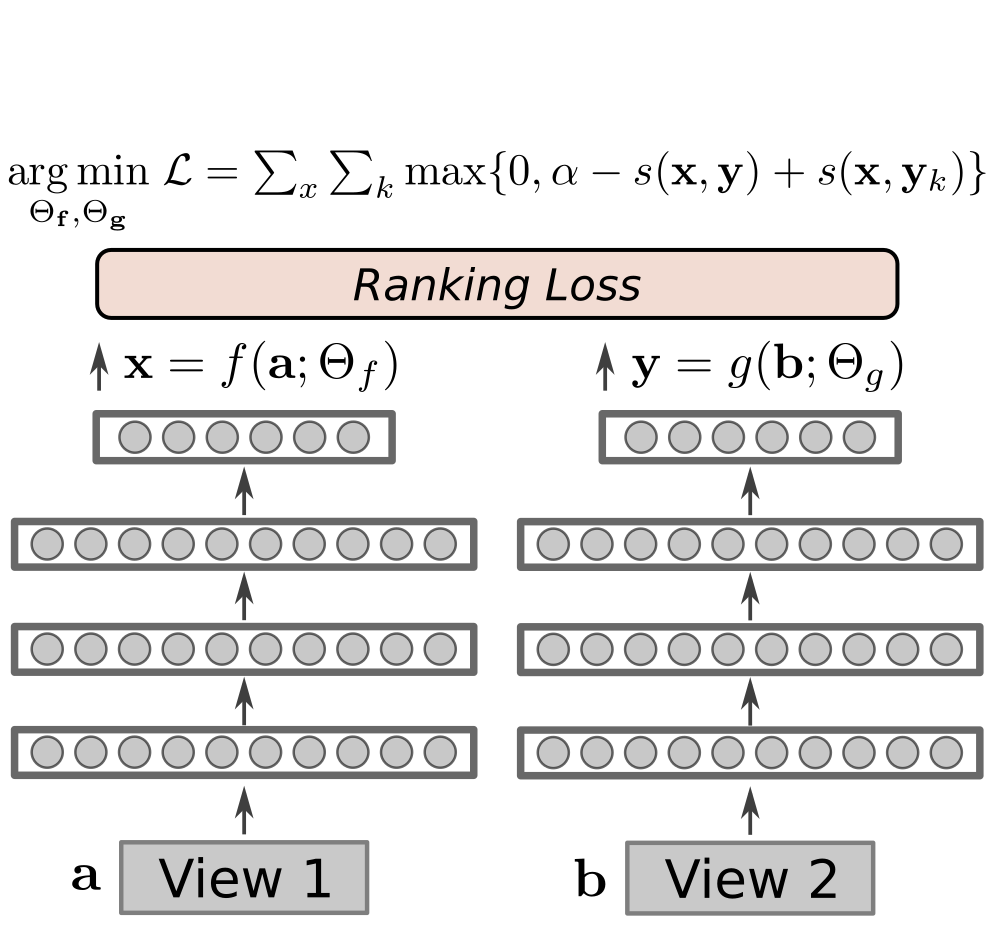} }}%
\quad
\subfloat[Canonically correlated projection layer optimized with ranking loss (\emph{CCAL-$\mathcal{L}_{rank}$}).]{\label{fig:ccal_sketch}{\includegraphics[height=0.27\textwidth]{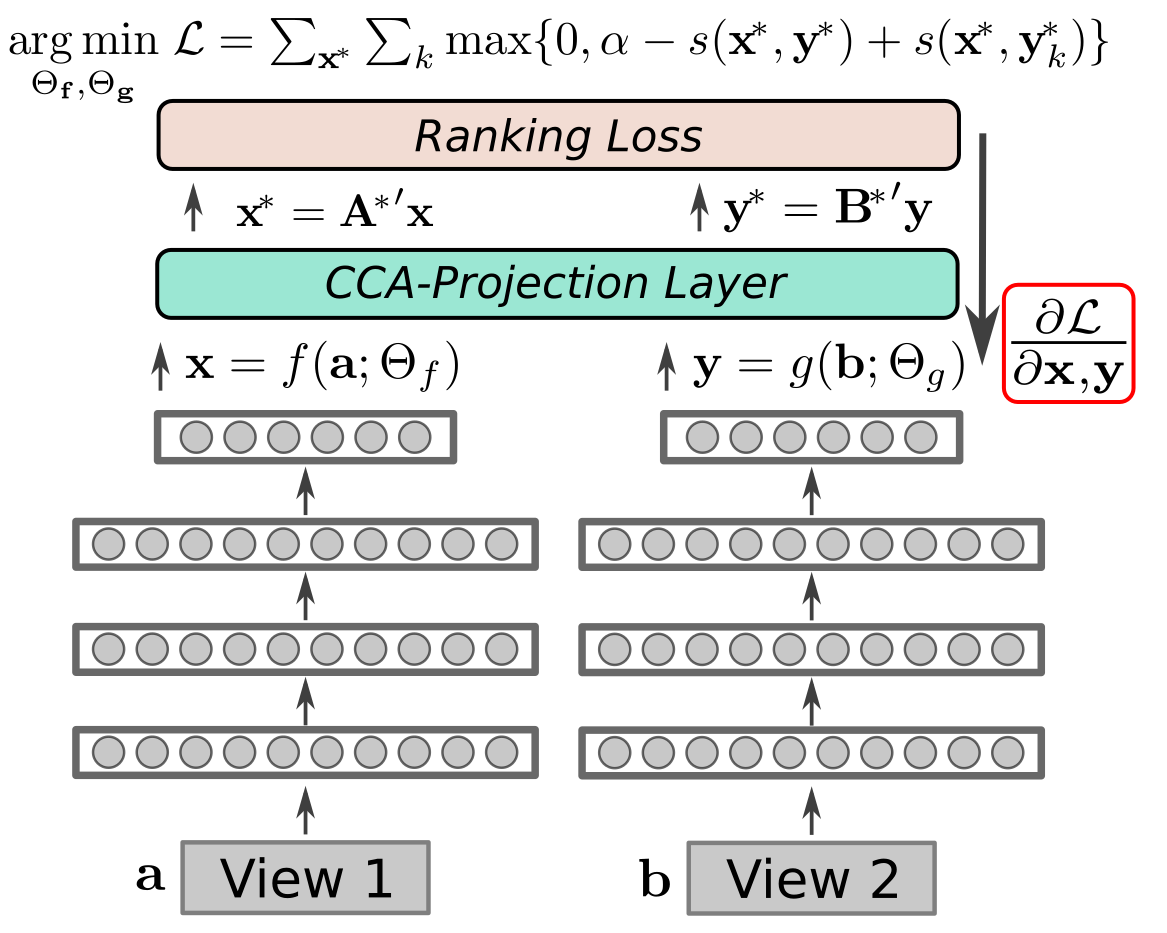} }}%
\caption{Sketches of cross-modality retrieval networks.
         The proposed model in (c) unifies (a) and (b) and takes advantage of both: componentwise correlated CCA projections
         and a pairwise ranking loss for cross-modality embedding space learning.
         We emphasize that our proposal in (c) requires to backpropagate the ranking loss $\mathcal{L}$ through the analytical computation of the
         optimally correlated CCA embedding projections $\mAstar$ and $\mBstar$ (see Equation (\ref{eq:cca_solution})).
         We thus need to compute their partial derivatives with
         respect to the network's hidden representations $\vx$ and $\vy$, i.e.\ 
         $\frac{\partial \mAstar}{\partial \vx, \vy}$ and $\frac{\partial \mBstar}{\partial \vx, \vy}$ (addressed in Section \ref{sec:ccal}).
\label{fig:comparison_lda_cce}
}
\end{figure*}

Figure~\ref{fig:comparison_lda_cce} compares our proposed approach to the alternatives discussed above.
DCCA defines an objective optimizing a dual-view neural network such that its two views will be maximally correlated (Figure~\ref{fig:dcca_sketch}).
Pairwise ranking losses are loss functions to optimize a dual-view neural network such that its two views are well-suited for nearest-neighbor retrieval in the embedding space (Figure~\ref{fig:learned_sketch}).
In our approach, we boost optimization of a pairwise ranking loss based on cosine distance by placing a special-purpose layer, the CCA projection layer, between a dual-view neural network and the optimization target (Figure~\ref{fig:ccal_sketch}).
Our experiments in Section \ref{sec:experiments} will show the effectiveness of this proposal.

\section{Canonical Correlation\\ Analysis}
\label{sec:methods_cca}
%
%

\label{sec:classic_cca}
In this section we review the concepts of CCA, the basis for our methodology. Let
$\vx \in \mathbb{R}^{d_x}$
and
$\vy \in \mathbb{R}^{d_y}$
denote two random column vectors with covariances $\Sigma_{xx}$ and $\Sigma_{yy}$ and cross-covariance $\Sigma_{xy}$.
The objective of CCA is to find two matrices $\mAstar \!\in \mathbb{R}^{d_x \times k}$ and $\mBstar \!\in \mathbb{R}^{d_y \times k}$ 
composed of $k$ paired column vectors $\mA_j$ and $\mB_j$
(with $k \leq d_x$ and $k \leq d_y$) that project $\vx$ and $\vy$ into a common space maximizing their componentwise correlation:
\begin{align}
(\mAstar, \mBstar) &= \argmax_{\mA, \mB}\sum_{j=1}^{k}~\corr(\mA_j'\vx, \mB_j'\vy) \\
                   &= \argmax_{\mA, \mB}~\sum_{j=1}^{k}~\frac{\mA_j'\Sxy\mB_j}{\sqrt{\mA_j'\Sxx\mA_j \; \mB_j'\Syy\mB_j}}
\end{align}
Since the objective of CCA is invariant to scaling of the projection matrices, we constrain the projected dimensions to have unit variance. Furthermore, CCA seeks subsequently uncorrelated projection vectors, arriving at the equivalent formulation:
\begin{align}
(\mAstar, \mBstar) 
                   &= \argmax_{\mA'\Sxx\mA = \mB'\Syy\mB = \mathbf{I}_k}~\trace\left(\mA'\Sxy\mB\right)
\label{eq:cca_obj}
\end{align}
Let $\mT = \Sxx^{-1/2} \Sxy \Syy^{-1/2}$, and let $\mU\diag(\vd)\mV'$ be the Singular Value Decomposition (SVD) of $\mT$ with ordered singular values $d_i \geq d_{i+1}$.
As shown in \cite{Mardia1979}, we obtain $\mAstar$ and $\mBstar$ from the top $k$ left- and right-singular vectors of $\mT$:
\begin{equation}
\mAstar = \Sxx^{-1/2} \mU_{:k}
\qquad
\mBstar = \Syy^{-1/2} \mV_{:k}
\label{eq:cca_solution}
\end{equation}
Moreover, the correlation in the projection space is the sum of the top $k$ singular values:%
\footnote{We understand the correlation of two vectors to be defined as $\corr(\vx, \vy) = \sum_{i}\sum_{j} \corr(x_i, y_j)$.}
\begin{equation}
\corr({\mAstar}' \vx, {\mBstar}' \vy) = \sum_{i \leq k} d_i
\label{eq:cca_sv}
\end{equation}
In practice, the covariances and cross-covariance of $\vx$ and $\vy$ are usually not known,
but estimated from a training set of $m$ paired vectors,
expressed as matrices $\mX \in \mathbb{R}^{d_x \times m}, \mY \in \mathbb{R}^{d_y \times m}$ by:
\begin{equation}
\Shxx = \frac{1}{m-1} \overbar\mX \overbar\mX' + r \mathbf{I}
\; \text{ and } \; \Shxy = \frac{1}{m-1} \overbar\mX \overbar\mY'.
\end{equation}
$\overbar\mX$ is the centered version of $\mX$. $\Shyy$ is defined analogously to $\Shxx$.
Additionally, we apply a regularization parameter $r \mathbf{I}$ to ensure that the covariance matrices are positive definite.
Substituting these estimates for $\Sxx$, $\Sxy$ and $\Syy$, respectively, we can compute $\mAstar$ and $\mBstar$ using Equation~(\ref{eq:cca_solution}).

\section{Cross-Modality Retrieval\\ Baselines}
\label{sec:xm_retrieval}
In this section we review the two most related works forming the basis for our approach.

\subsection{Deep Canonical Correlation\\ Analysis}
Andrew et al. \cite{Andrew2013DCCA} propose an extension of CCA to learn parametric nonlinear transformations of two random vectors, such that their correlation is maximized. Let $\va \in \mathbb{R}^{d_a}$ and $\vb \in \mathbb{R}^{d_b}$ denote two random vectors, and let $\vx = f(\va; \Theta_f)$ and $\vy = g(\vb; \Theta_g)$ denote their nonlinear transformations, parameterized by $\Theta_f$ and $\Theta_g$.
DCCA optimizes the parameters $\Theta_f$ and $\Theta_g$ to maximize the correlation of the topmost hidden representations $\vx$ and $\vy$.
For $d_x = d_y = k$, this objective corresponds to Equation~\ref{eq:cca_sv}, i.e., the sum of all singular values of $\mT$, also called the trace norm:
\begin{equation}
\corr({\mAstar}'f(\va; \Theta_f), {\mBstar}'g(\vb; \Theta_g)) = ||\mT||_{\text{tr}} \text{.}
\label{eq:tno}
\end{equation}
Andrew et al. \cite{Andrew2013DCCA} show how to compute the gradient of this \emph{Trace Norm Objective} (TNO) with respect to $\vx$ and $\vy$.
Assuming $f$ and $g$ are differentiable with respect to $\Theta_f$ and $\Theta_g$ (as is the case for neural networks), this allows to optimize the nonlinear transformations via gradient-based methods.

\begin{figure}[ht]
 \centerline{\includegraphics[width=0.8\columnwidth]{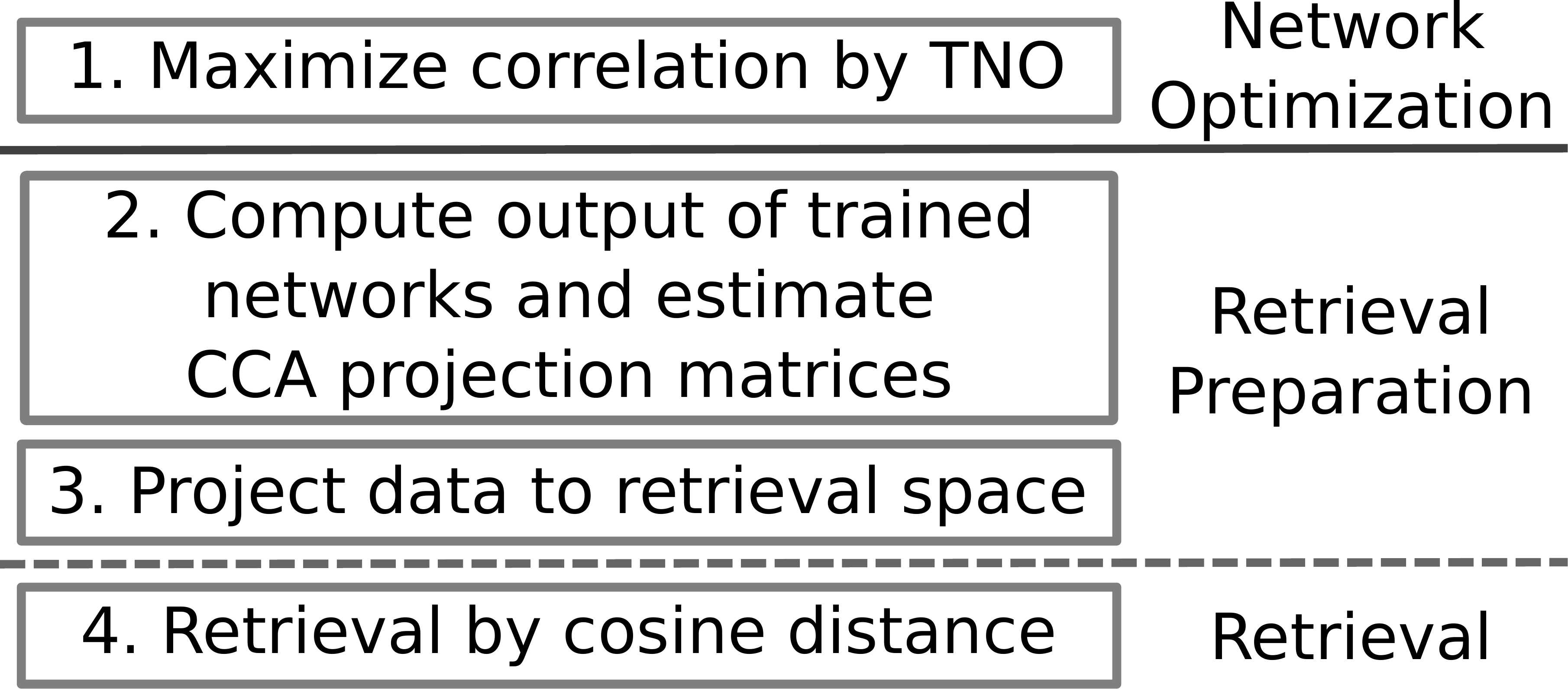}}
 \caption{\emph{DCCA} retrieval pipeline proposed in \cite{Yan2015DeepCorr}.
          Note that all processing steps below the solid line are performed after network optimization is complete.}
 \label{fig:e2e_dcca}
\end{figure}

Yan and Mikolajczyk \cite{Yan2015DeepCorr} suggest the following procedure to utilize DCCA for cross-modality retrieval:
first, neural networks $f$ and $g$ are trained using the TNO, with $\va$ and $\vb$ representing different views of an entity (e.g.\ image and text); then, after the training is finished, the CCA projections are computed using Equation~(\ref{eq:cca_solution}), and all retrieval candidates are projected into the embedding space; finally, at test time, queries of either modality are  projected into the embedding space, and the best-matching sample from the other modality is found through nearest neighbor search using the cosine distance.
Figure \ref{fig:e2e_dcca} provides a summary of the entire retrieval pipeline.
In our experiments, we will refer to this approach as \emph{\DCCA}.

DCCA is limited by design to use the objective function described in Equation \eqref{eq:tno}, and only seeks to maximize the correlation in the embedding space. During training, the CCA projection matrices are never computed, nor are the samples projected into the common retrieval space. All the retrieval steps---most importantly, the computation of CCA projections---are performed only once after the networks $f$ and $g$ have been optimized. This restricts potential applications, because we cannot use the projected data as an input to subsequent layers or task-specific objectives. We will show how our approach overcomes this limitation in Section \ref{sec:ccal}.

\subsection{Pairwise Ranking Loss}

Kiros et al. \cite{kiros2014unifying} learn a multi-modal joint embedding space for images and text. They use the cosine of the angle between two corresponding vectors $\mathbf{x}$ and $\mathbf{y}$ as a scoring function, i.e., $s(\mathbf{x}, \mathbf{y}) = \cos(\mathbf{x}, \mathbf{y})$. Then, they optimize a pairwise ranking loss
\begin{equation}
\mathcal{L}_{rank}=\sum_{\mathbf{x}} \sum_{k} \max \{0, \alpha - s(\mathbf{x}, \mathbf{y}) + s(\mathbf{x}, \mathbf{y}_k) \}
\label{eq:contrastive}
\end{equation}
where $\mathbf{x}$ is an embedded sample of the first modality,
$\mathbf{y}$ is the matching embedded sample of the second modality,
and $\mathbf{y}_k$ are the contrastive (mismatching) embedded samples of the second modality
(in practice, all mismatching samples in the current mini-batch).
The hyper-parameter $\alpha$ defines the margin of the loss function.
This loss encourages an embedding space where the cosine distance between matching samples is lower
than the cosine distance of mismatching samples.

In this setting, the networks $f$ and $g$ have to learn the embedding projections freely from randomly initialized weights.
Since the projections are learned from scratch by optimizing a ranking loss, in our experiments we denote this approach by \emph{Learned-$\mathcal{L}_{rank}$}. Figure \ref{fig:learned_sketch} shows a sketch of this paradigm.

\section{Learning with Canonically Correlated Embedding Projections}
\label{sec:ccal}
In the following we explain how to bring both concepts --- DCCA and Pairwise Ranking Losses --- together
to enhance cross-modality embedding space learning.
\subsection{Motivation}
\label{subsec:ccal_motivation}
We start by providing an intuition on why we expect this combination to be fruitful:
\emph{\DCCA} maximizes the correlation between the latent representations of two different neural networks
via the TNO derived from classic CCA.
As correlation and cosine distance are related, we can also use such a network for cross-modality retrieval \cite{Yan2015DeepCorr}.
Kiros et al. \cite{kiros2014unifying}, on the other hand, learn a cross-modality retrieval embedding by optimizing
an objective customized for the task at hand.
The motivation for our approach is that we want to benefit from both: a task specific retrieval objective,
and componentwise optimally correlated embedding projections.

To achieve this, we devise a \emph{CCA layer} that analytically computes the CCA projections $\mAstar$ and $\mBstar$ during training, and projects incoming samples into the embedding space. The projected samples can then be used in subsequent layers, or for computing task-specific losses such as the pairwise ranking loss. 
Figure~\ref{fig:ccal_sketch} illustrates the central idea of our combined approach.
Compared to Figure \ref{fig:learned_sketch}, we insert an additional linear transformation.
However, this transformation is not learned (otherwise it could be merged with the previous layer, which is not followed by a nonlinearity).
Instead, it is computed to be the transformation that maximizes componentwise correlation between the two views.
$\mAstar$ and $\mBstar$ in Figure~\ref{fig:ccal_sketch} are the very projections given by Equation (\ref{eq:cca_solution}) in Section \ref{sec:classic_cca}.

In theory, optimizing a pairwise ranking loss alone could yield projections equivalent to the ones computed by CCA.
In practice, however, we observe that the proposed combination gives much better cross-modality retrieval results (see Section \ref{sec:experiments}).

Our design requires backpropagating errors through the analytical computation of the CCA projection matrices.
DCCA \cite{Andrew2013DCCA} does not cover this, since projecting the data is not necessary for optimizing the TNO.
In the remainder of this section, we discuss how to establish gradient flow (backpropagation)
through CCA's optimal projection matrices.
In particular, we require the partial derivatives
$\frac{\partial \mAstar}{\partial \vx, \vy}$ and $\frac{\partial \mBstar}{\partial \vx, \vy}$
of the projections with respect to their input representations $\vx$ and $\vy$.
This will allow us to use CCA as a layer within a multi-modality neural network,
instead of as a final objective (TNO) for correlation maximization only.

\subsection{Gradient of CCA Projections}
\label{subsec:grad_of_cca}
As mentioned above, we can compute the canonical correlation along with the optimal projection matrices from the singular value decomposition $\mT = \Sxx^{-1/2} \Sxy \Syy^{-1/2} = \mU \diag(\vd) \mV'$.
Specifically, we obtain the correlation as $\corr({\mAstar}' \vx, {\mBstar}' \vy) = \sum_i d_i$, and the projections as $\mAstar = \Sxx^{-1/2} \mU$ and $\mBstar = \Syy^{-1/2} \mV$.
For DCCA, it suffices to compute the gradient of the total correlation wrt.\ $\vx$ and $\vy$ in order to backpropagate it through the two networks $f$ and $g$.
Using the chain rule, Andrew et al. decompose this into the gradients of the total correlation wrt.\ $\Sxx$, $\Sxy$ and $\Syy$, and the gradients of those wrt.\ $\vx$ and $\vy$ \cite{Andrew2013DCCA}.
Their derivations of the former make use of the fact that both the gradient of $\sum_i d_i$ wrt.\ $\mT$ and the gradient of $||\mT||_{\text{tr}}$ (the trace norm objective in Equation~(\ref{eq:tno})) wrt.\ $\mT'\mT$ have a simple form; see Section 7 in \cite{Andrew2013DCCA} for details.

In our case where we would like to backpropagate errors through the CCA transformations,
we instead need the gradients of the projected data $\vx^{\!*}={\mAstar}' \vx$ and $\vy^{\!*}={\mBstar}' \vy$ wrt.\ $\vx$ and $\vy$,
which requires the partial derivatives $\frac{\partial \mAstar}{\partial \vx, \vy}$ and $\frac{\partial \mBstar}{\partial \vx, \vy}$.
We could again decompose this into the gradients wrt.\ $\mT$, the gradients of $\mT$ wrt.\ $\Sxx$, $\Sxy$ and $\Syy$ and the gradients of those wrt.\ $\vx$ and $\vy$.
However, while the gradients of $\mU$ and $\mV$ wrt.\ $\mT$ are known \cite{Papadopoulo2000_svdgrad}, they involve solving $O((d_x d_y)^2)$ linear $2\!\times\!2$ systems.
Instead, we reformulate the solution to use two symmetric eigendecompositions $\mT\mT' = \mU \diag(\ve) \mU'$ and $\mT'\mT = \mV \diag(\ve) \mV'$ (Equation~270 in \cite{Cookbook}).
This gives us the same left and right eigenvectors we would obtain from the SVD,
along with the squared singular values ($e_i = d_i^2$).
The gradients of eigenvectors of symmetric real eigensystems have a simple form \cite{Magnus1985_eighgrad} and both $\mT\mT'$ and $\mT'\mT$ are differentiable wrt.\ $\vx$ and $\vy$.

To summarize: in order to obtain an efficiently computable definition
of the gradient for CCA projections, we have reformulated
the forward pass (the computation of the CCA transformations).
Our formulation using two eigen-decompositions translates into a series of
computation steps that are differentiable in a graph-based,
auto-differentiating math compiler such as \emph{Theano} \cite{Theano}, which, together
with the chain rule, gives an efficient implementation of the
CCA layer gradient for training our network\footnote{The code of our implementation of the CCA layer is available at \repo .}.
For a detailed description of the CCA layer forward pass we refer to Algorithm \ref{nips2017:algo:cca_layer} in the Appendix of this article.
As the technical implementation is not straight-forward, we also discuss the crucial steps in the Appendix.

Thus,
we now have the means to benefit from the optimal CCA projections
but still optimize for a task-specific objective.
In particular, we utilize the \emph{pairwise ranking loss} of Equation (\ref{eq:contrastive})
on top of an intermediate CCA embedding projection layer.
We denote the proposed retrieval network of Figure~\ref{fig:ccal_sketch} as \emph{CCAL-$\mathcal{L}_{rank}$} in our experiments (\emph{CCAL} refers to CCA Layer).

\section{Experiments}
\label{sec:experiments}
We evaluate our approach (\emph{CCAL-$\mathcal{L}_{rank}$}) in cross-modality retrieval experiments on
two image-to-text and one audio-to-sheet-music dataset.
Additionally, we provide results on two zero-shot text-to-image retrieval scenarios proposed in \cite{Reed_2016_VisualDescriptions}.
For comparison, we consider the approach of \cite{Yan2015DeepCorr} (\emph{\DCCA}),
our own implementation of the TNO (denoted by \emph{DCCA}),
as well as the freely learned projection embeddings (\emph{Learned-$\mathcal{L}_{rank}$}) optimizing the ranking loss of \cite{kiros2014unifying}.

\begin{figure}[ht!]
 \centerline{\includegraphics[width=0.9\columnwidth]{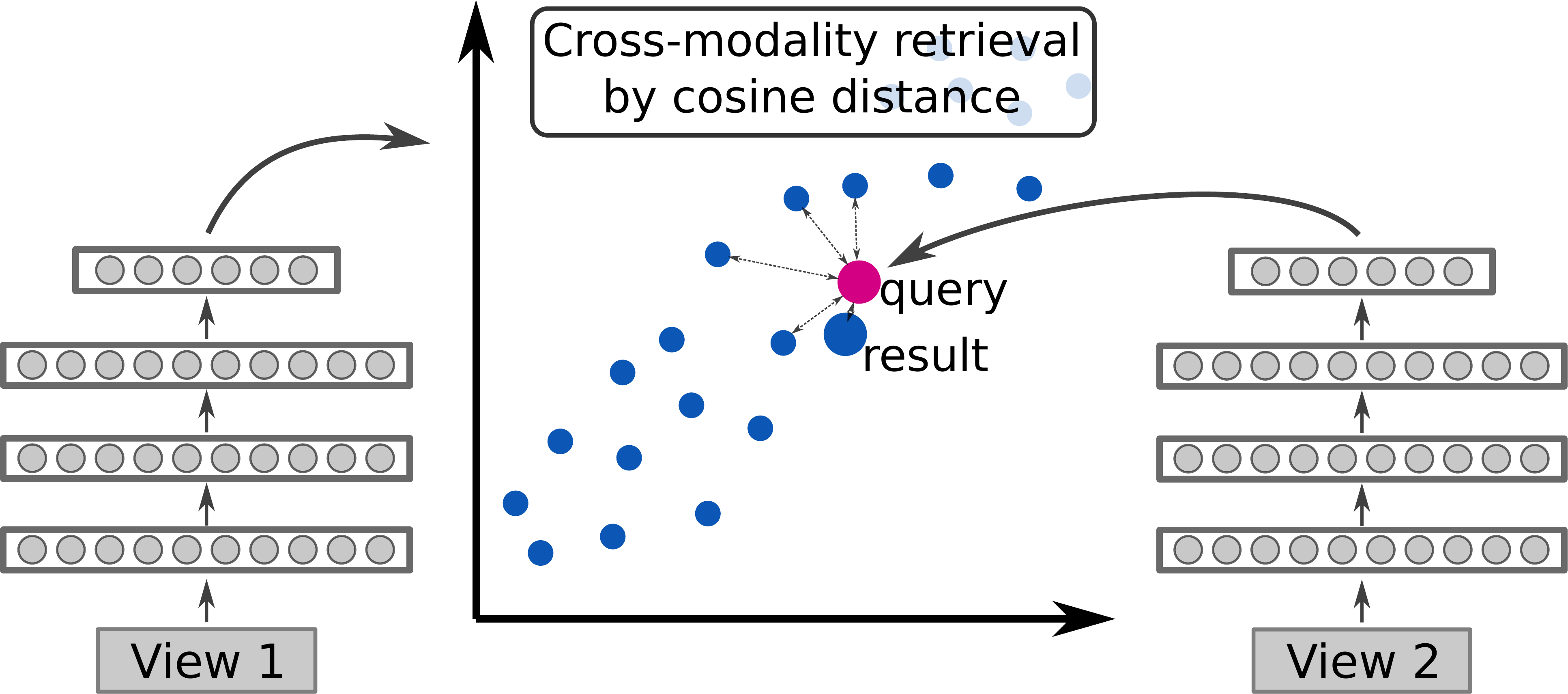}}
 \caption{Sketch of cross-modality retrieval.
 		  The blue dots are the embedded candidate samples.
 		  The red dot is the embedding of the search query.
          The larger blue dot highlights the closest candidate selected as the retrieval result.}
 \label{fig:nn_retrieval}
\end{figure}

The task for all three datasets is to retrieve the correct counterpart when given an instance of the other modality as a search query.
For retrieval, we use the cosine distance in embedding space for all approaches.
First, we embed all candidate samples of the target modality into the retrieval embedding space.
Then, we embed the query element $\mathbf{y}$ with the second network
and select its nearest neighbor $\mathbf{x}_j$ of the target modality.
Fig. \ref{fig:nn_retrieval} shows a sketch of this retrieval by embedding space learning paradigm.

As evaluation measures, we consider the \emph{Recall@k (R@k in \%)}
as well as the \emph{Median Rank (MR)} and the \emph{Mean Reciprocal Rank (MRR in \%)}.
The \emph{R@k} rate (higher is better) is the ratio of queries which have the correct corresponding counterpart in the first $k$ retrieval results.
The \emph{MR} (lower is better) is the median position of the target in a similarity-ordered list of available candidates.
Finally, we define the \emph{MRR} (higher is better) as the mean value of $1 / rank$ over all queries
where $rank$ is again the position of the target in the similarity ordered list of available candidates.

\subsection{Image-Text Retrieval}
\label{subsec:imge-text}
In the first part of our experiments, we consider \emph{Flickr30k} and \emph{IAPR TC-12},
two publicly available datasets for image-text cross-modality retrieval.
Flickr30k consists of image-caption pairs,
where each image is annotated with five different textual descriptions.
The train-validation-test split for Flickr30k is 28000-1000-1000.
In terms of evaluation setup, we follow \emph{Protocol 3} of \cite{Yan2015DeepCorr}
and concatenate the five available captions into one,
meaning that only one, but richer text annotation remains per image.
This is done for all three sets of the split.
The second image-text dataset, IAPR TC-12, contains 20000 natural images where only one---but compared to Flickr30k more detailed---caption
is available for each image.
As no predefined train-validation-test split is provided,
we randomly select 1000 images for validation and 2000 for testing, and keep the rest for training.
\cite{Yan2015DeepCorr} also use 2000 images for testing, but did not explicitly mention holdout images for validation.
Table \ref{tab:example_images} shows an example image along with its corresponding captions or caption for either dataset.

\begin{table*}[t!]
  \centering
  \begin{tabular}{cm{10.0cm}}
    \begin{minipage}{.17\textwidth}
    		\includegraphics[width=0.75\textwidth]{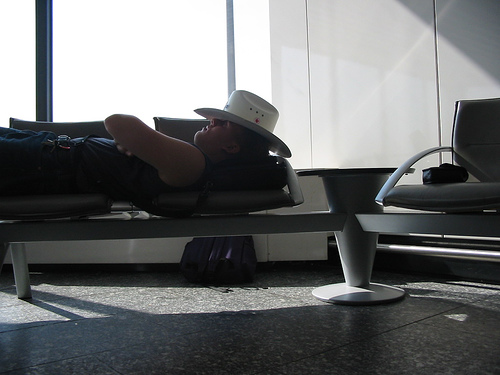}
    \end{minipage}
    &
      \small
      \begin{itemize}[wide, labelwidth=!, labelindent=0pt]
      	\itemsep -0.2em
        \item [] A man in a white cowboy hat reclines in front of a window in an airport.
        \item [] A young man rests on an airport seat with a cowboy hat over his face.
        \item [] A woman relaxes on a couch , with a white cowboy hat over her head.
        \item [] A man is sleeping inside on a bench with his hat over his eyes.
        \item [] A person is sleeping at an airport with a hat on their head.
      \end{itemize}
    \\
    \begin{minipage}{.17\textwidth}
    		\includegraphics[width=0.75\textwidth]{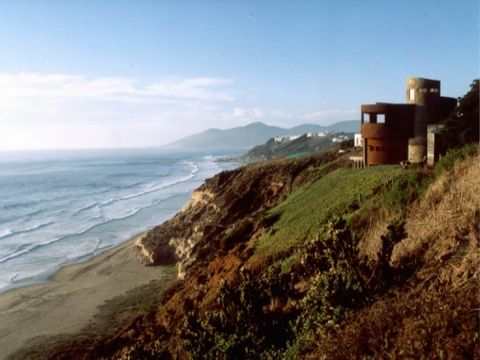}
    \end{minipage}
    &
    \small
      \begin{itemize}[wide, labelwidth=!, labelindent=0pt]
        \item [] a green and brown embankment with brown houses on the right and a light brown sandy beach at the dark blue sea on the left; a dark mountain range behind it and white clouds in a light blue sky in the background;
      \end{itemize}
  \end{tabular}
  \caption{Example images for Flickr30k (top) and IAPR TC-12 (bottom)}\label{tab:example_images}
\end{table*}

The input to our networks is a $4096$-dimensional image feature vector
along with a corresponding text vector representation
which has dimensionality $5793$ for Flickr30k and $2048$ for IAPR TC-12.
The image embedding is computed from the last hidden layer of a network pretrained
on ImageNet \cite{imagenet_cvpr09} (layer \emph{fc7} of \emph{CNN\_S} by \cite{Chatfield14}).
In terms of text pre-processing, we follow \cite{Yan2015DeepCorr},
tokenizing and lemmatizing the raw captions as the first step.
Based on the lemmatized captions, we compute $l2$-normalized TF/IDF-vectors,
omitting words with an overall occurrence smaller than five for Flickr30k and three for IAPR TC-12, respectively.
The image representation is processed by a linear dense layer
with $128$ units, which will also be the dimensionality $k$ of the resulting retrieval embedding.
The text vector is fed through two batch-normalized \cite{LoffeS2015BatchNorm}
dense layers of $1024$ units each and the ELU activation function \cite{Clevert2015ELU}.
As a last layer for the text representation network, we again apply
a dense layer with $128$ linear units.

For a fair comparison, we keep the structure and number of parameters of all networks in our experiments the same.
The only difference between the networks are the objectives
and the hyper-parameters used for optimization.
Optimization is performed using Stochastic Gradient Descent (SGD) with
the \emph{adam} update rule \cite{Kingma2014adam} (for details please see our appendix).

Table \ref{tab:result-iapr} lists our results on IAPR TC-12.
Along with our experiments, we also show the results reported in \cite{Yan2015DeepCorr}
as a reference (\emph{\DCCA}).
However, a direct comparison to our results may not be fair:
\emph{\DCCA} uses a different ImageNet-pretrained network for the image representation,
and finetunes this network while we keep it fixed.
This is because our interest is in comparing the methods in a stable setting, not in obtaining the best possible results.
Our implementation of the TNO (\emph{DCCA}) uses the same objective as \emph{\DCCA},
but is trained using the same network architecture as our remaining models and permits a direct comparison.
Additionally, we repeat each of the experiments 10 times with different initializations
and report the mean for each of the evaluation measures.
\begin{table*}[ht!]
  \caption{Retrieval results on IAPR TC-12.
           ``\DCCA'' is taken from \cite{Yan2015DeepCorr}.}
  \label{tab:result-iapr}
  \centering
  \begin{tabular}{lcccccccccc}
    \toprule
    & \multicolumn{5}{c}{Image-to-Text} & \multicolumn{5}{c}{Text-to-Image} \\
	\midrule    
    Method				& R@1	& R@5	& R@10	& MR	& MRR & R@1	& R@5	& R@10	& MR	& MRR \\
    \midrule
    \DCCA   							& 30.2	& 57.0	& -		& - 	& 42.6 & 29.5	& 60.0	& -		& -		& 41.5 \\
    DCCA   								& 31.0 	& 58.7 	& 70.4 	& 3.6 	& 43.9 & 29.5 	& 58.2 	& 70.5 	& 4.0	& 42.7 \\
	Learned-$\mathcal{L}_{rank}$		& 22.3  & 50.7 	& 63.8 	& 5.2	& 35.7 & 21.6 	& 50.1 	& 63.3 & 5.5	& 35.1 \\
    CCAL-$\mathcal{L}_{rank}$   		& 31.6 	& 61.0  & 72.2 	& 3.0	& 45.0 & 29.6 	& 60.0 	& 72.2 	& 3.6	& 43.5 \\ 	   
    \bottomrule
  \end{tabular}
\end{table*}

When taking a closer look at Table \ref{tab:result-iapr},
we observe that our results achieved by optimizing the TNO (\emph{DCCA})
surpass the results reported in \cite{Yan2015DeepCorr}.
We already discussed above that the two versions are not directly comparable.
However, given this result,
we consider our implementation of \emph{DCCA} as a valid baseline for our experiments in Section \ref{subsec:audio_sheet}
where no results are available in the literature.
When looking at the performance of \emph{CCAL-$\mathcal{L}_{rank}$}
we further observe that it outperforms all other methods,
although the difference to \emph{DCCA} is not pronounced for all of the measures.
Comparing \emph{CCAL-$\mathcal{L}_{rank}$} with the freely-learned projection matrices (\emph{Learned-$\mathcal{L}_{rank}$})
we observe a much larger performance gap.
This is interesting, as in principle the learned projections could converge to exactly the same solution as \emph{CCAL-$\mathcal{L}_{rank}$}.
We take this as a quantitative confirmation
that the learning process benefits from CCA's optimal projection matrices.

In Table \ref{tab:result-flickr30k}, we list our results on the Flickr30k dataset.
As above, we show the retrieval performances of \cite{Yan2015DeepCorr} as a baseline along with our results
and observe similar behavior as on IAPR TC-12.
Again, we point out the poor performance of the freely-learned projections (\emph{Learned-$\mathcal{L}_{rank}$}) in this experiment.
Keeping this observation in mind, we will notice a different behavior
in the experiments in Section \ref{subsec:audio_sheet}.

Note that there are various other methods reporting results on Flickr30k \cite{karpathy2014deep,socher2014grounded,mao2014explain,kiros2014unifying} which partly surpass ours, for example by using more elaborate processing of the textual descriptions or more powerful ImageNet models.
We omit these results as we focus on
the comparison of \emph{DCCA} and freely-learned projections with the proposed CCA embedding layer.
\begin{table*}[ht!]
  \caption{Retrieval results on Flickr30k.
  ``\DCCA'' is taken from \cite{Yan2015DeepCorr}.}
  \label{tab:result-flickr30k}
  \centering
  \begin{tabular}{lcccccccccc}
    \toprule
    & \multicolumn{5}{c}{Image-to-Text} & \multicolumn{5}{c}{Text-to-Image} \\
	\midrule    
    Method				& R@1	& R@5	& R@10	& MR	& MRR & R@1	& R@5	& R@10	& MR	& MRR \\
    \midrule
    \DCCA   							& 27.9	& 56.9	& 68.2	& 4		& -    & 26.8	& 52.9	& 66.9	& 4		& - \\
    DCCA   								& 31.6 	& 59.2 	& 69.3 	& 3.3	& 44.2 & 30.3 	& 58.3 	& 69.2 	& 3.8	& 43.1 \\
	Learned-$\mathcal{L}_{rank}$		& 23.7 	& 50.5 	& 63.0 	& 5.3	& 36.3 & 23.6 	& 51.0 	& 62.5 	& 5.2	& 36.5 \\
    CCAL-$\mathcal{L}_{rank}$   		& 32.0 	& 59.2 	& 70.4 	& 3.2	& 44.8 & 29.9 	& 58.8 	& 70.2 	& 3.7	& 43.3 \\
    \bottomrule
  \end{tabular}
\end{table*}

\subsection{Audio-Sheet-Music Retrieval}
\label{subsec:audio_sheet}
\begin{figure}[ht]
 \centerline{\includegraphics[width=0.99\columnwidth]{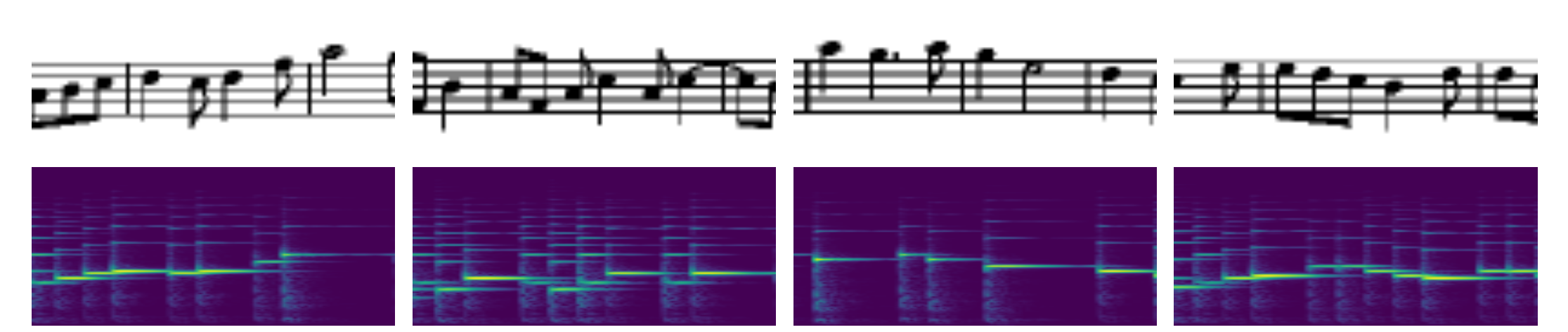}}
 \caption{Example of the data considered for audio-sheet-music (image) retrieval.
 		   Top: short snippets of sheet music images.
 		   Bottom: Spectrogram excerpts of the corresponding music audio.}
 \label{fig:correspondence_samples}
\end{figure}

For the second set of experiments, we consider the Nottingham piano midi dataset \cite{Boulanger_2012_ModSequ}.
The dataset is a collection of midi files split into train, validation and test set
already used by \cite{Dorfer2016Towards} for experiments on end-to-end score-following in sheet-music images.
Here, we tackle the problem of audio-sheet-music retrieval, i.e.
matching short snippets of music (audio) to corresponding parts in the sheet music (image).
Figure \ref{fig:correspondence_samples} shows examples of such correspondences.

We conduct this experiment for two reasons:
First, to show the advantage of the proposed method over different domains.
Second, the data and application is of high practical relevance in the domain of Music Information Retrieval (MIR).
A system capable of linking sheet music (images) and the corresponding music (audio)
would be useful in many content-based musical retrieval scenarios.

In terms of audio preparation, we compute log frequency spectrograms with a sample rate of 22.05kHz,
a FFT window size of 2048, and a computation rate of 31.25 frames per second.
These spectrograms (136 frequency bins) are then directly fed into
the audio part of the cross-modality networks.
Figure \ref{fig:correspondence_samples} shows a set of audio-to-sheet correspondences
presented to our network for training.
One audio excerpt comprises $100$ frames and the dimension
of the sheet image snippet is $40 \times 100$ pixels.
Overall this results in $270,\!705$ train, $18,\!046$ validation and $16,\!042$ test audio-sheet-music pairs.
This is an order of magnitude more training data than for the image-to-text datasets of the previous section.

In the experiments in Section \ref{subsec:imge-text}, we relied on pre-trained ImageNet features
and relatively shallow fully connected text-feature processing networks. 
The model here differs from this, as it consists of two deep convolutional networks learned entirely from scratch.
Our architecture is a VGG-style \cite{simonyan2014very} network consisting of sequences of $3 \times 3$ convolution stacks followed by $2 \times 2$ max pooling.
To reduce the dimensionality to the desired correlation space dimensionality $k$ (in this case 32)
we insert as a final building block a $1 \times 1$ convolution having $k$ feature maps
followed by global average pooling \cite{LinCY2013NIN} (for further architectural details we again refer to the appendix of this manuscript).

Table \ref{tab:result-audio2sheet} lists our result on audio-to-sheet music retrieval.
As in the experiments on images and text, the proposed CCA projection embedding layer trained
with pairwise ranking loss outperforms the other models.
Recalling the results from Section \ref{subsec:imge-text}, we observe an increased performance of
the freely-learned embedding projections.
On measures such as R@5 or R@10 it achieves similar to or better performance than \emph{DCCA}.
One of the reasons for this could be the fact that there is an order of magnitude more training data available for
this task to learn the projection embedding from random initialization.
Still, our proposed combination of both concepts (\emph{CCAL-$\mathcal{L}_{rank}$}) achieves highest retrieval scores.
\begin{table*}[ht!]
  \caption{Retrieval results on Nottingham dataset (Audio-to-Sheet-Music Retrieval).}
  \label{tab:result-audio2sheet}
  \centering
  \begin{tabular}{lcccccccccc}
    \toprule
    & \multicolumn{5}{c}{Sheet-to-Audio} & \multicolumn{5}{c}{Audio-to-Sheet} \\
	\midrule    
    Method				& R@1	& R@5	& R@10	& MR	& MRR & R@1	& R@5	& R@10	& MR	& MRR \\
    \midrule
    DCCA   								& 42.0	& 88.2 	& 93.3 & 2	& 62.2 & 44.6	& 87.9 	& 93.2	& 2	& 63.5 \\
	Learned-$\mathcal{L}_{rank}$		& 40.7	& 89.6 	& 95.6 & 2	& 61.7 & 41.4	& 88.9 	& 95.4	& 2	& 61.9 \\
    CCAL-$\mathcal{L}_{rank}$   		& 44.1	& 93.3 	& 97.7 & 2	& 65.3 & 44.5	& 91.6 	& 96.7	& 2	& 64.9 \\
    \bottomrule
  \end{tabular}
\end{table*}

\subsection{Performance in\\ Small Data Regime}
\label{sec:discussion}
The above results suggest that the benefit of using a CCA projection layer (\emph{CCAL-$\mathcal{L}_{rank}$})
over a freely-learned projection becomes most evident when few training data is available.
To examine this assumption, we repeat the audio-to-sheet-music experiment of the previous section,
but use only $10\%$ of the original training data ($\approx 27000$ samples).
We stress the fact that the learned embedding projection of \emph{Learned-$\mathcal{L}_{rank}$}
could converge to exactly the same solution as the CCA projections of \emph{CCAL-$\mathcal{L}_{rank}$}.
Table \ref{tab:result-audio2sheet_27k} summarizes the low data regime results for the three methods.
Consistent with our hypothesis, we observe a larger gap between \emph{Learned-$\mathcal{L}_{rank}$} and \emph{CCAL-$\mathcal{L}_{rank}$}
compared to the one obtained with all training data in Table \ref{tab:result-audio2sheet}.
We conclude that a network might be able to learn suitable embedding projections when sufficient training data is available.
However, when having fewer training samples, the proposed CCA projection layer strongly supports embedding space learning.
In addition, we also looked into the retrieval performance of \emph{Learned-$\mathcal{L}_{rank}$} and \emph{CCAL-$\mathcal{L}_{rank}$} on the training set
and observe comparable performance.
This indicates that the CCA layer also acts as a regularizer and helps to generalize to unseen samples.
\begin{table*}[ht!]
  \small
  \caption{Retrieval results on audio-to-sheet-music retrieval
           when using only $10\%$ of the train data.}
  \label{tab:result-audio2sheet_27k}
  \centering
  \begin{tabular}{lcccccccccc}
    \toprule
    				& \multicolumn{5}{c}{Sheet-to-Audio} & \multicolumn{5}{c}{Audio-to-Sheet} \\
	\midrule    
    Method &  R@1	& R@5	& R@10	& MR	& MRR & R@1	& R@5	& R@10	& MR	& MRR \\
    \midrule
    DCCA					& 20.0	& 53.6 	& 65.4 & 5	& 35.3 & 22.7 	& 54.7 	& 65.8	& 4	& 37.3 \\
	Learned-$\mathcal{L}_{rank}$		& 11.3	& 35.2 	& 47.6 & 12	& 23.0 & 12.6	& 35.2 	& 47.2	& 12 	& 23.7 \\
    CCAL-$\mathcal{L}_{rank}$   		& 22.2	& 59.2 	& 70.7 & 4	& 38.8 & 25.0	& 59.3 	& 70.9	& 4	& 40.4 \\	
    \bottomrule
  \end{tabular}
\end{table*}

\subsection{Zero-Shot Image-Text Retrieval}
\label{subsec:zero-shot}
Our last set of experiments focuses on a slightly modified retrieval setting,
namely image-text \emph{zero-shot retrieval} \cite{Reed_2016_VisualDescriptions}.
Given a set of image-text pairs originating from $C$ different categories
the data is split into a class-disjoint training, validation and test sets having no categorical overlap.
This implies that at test time we aim to retrieve images from textual queries describing categories (semantic concepts) never seen before, neither for training, nor for validation.

\begin{figure}[ht]
 \centerline{\includegraphics[width=0.99\columnwidth]{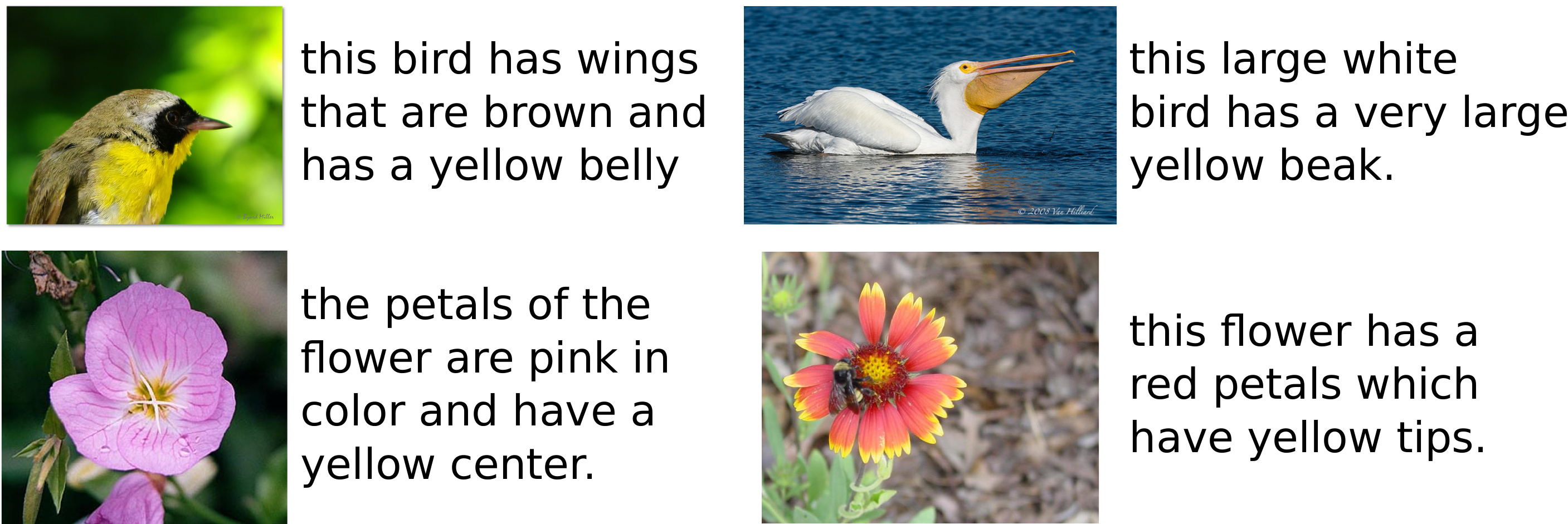}}
 \caption{Example images of CUB-200 birds and Oxford Flowers along with textual descriptions collected by Reed et al. \cite{Reed_2016_VisualDescriptions} for zero-shot retrieval from text.}
 \label{fig:cub_and_flowers}
\end{figure}

Reed et al. \cite{Reed_2016_VisualDescriptions} collected and provided textual descriptions for two publicly available datasets,
the \emph{CUB-200 bird image dataset} \cite{WelinderEtal2010} and the \emph{Oxford Flowers dataset} \cite{Nilsback08}.
According to the definition of zero-shot retrieval above we follow \cite{Reed_2016_VisualDescriptions}
and split CUB into 100 train, 50 validation and 50 test categories.
Flowers is split into 82 train and 20 validation / test classes respectively.
Figure \ref{fig:cub_and_flowers} shows some example images along with their textual descriptions.

Besides the modified, harder retrieval setting there is a second difference to the text-image retrieval experiments carried out in Section \ref{subsec:imge-text}.
Instead of using hand engineered textual features (e.g. TF-IDF) or unsupervised textual feature learning (e.g. word2vec \cite{mikolov2013distributed}) the authors in \cite{Reed_2016_VisualDescriptions} employ Convolutional Recurrent Neural Networks (CRNN) to learn the latent text representations directly from the raw descriptions.
In particular, they feed the descriptions as one-hot-word encodings to the text processing part of their networks.
In terms of image representations they still rely on $1024$-dimensional pretrained ImageNet features.
The feature learning part and the network architectures used for our experiments follows exactly the descriptions provided in \cite{Reed_2016_VisualDescriptions}.
The sole difference is, that we again replace the topmost embedding layer with the proposed CCA projection layer in combination with a pairwise ranking loss.

Table \ref{tab:result-zeroshot} compares the retrieval results of the respective methods on the two zero-shot retrieval datasets.
To allow for a direct comparison with the results reported in \cite{Reed_2016_VisualDescriptions} we follow their evaluation setup and report the \emph{Average Precision (AP@50)}.
The AP@50 is the percentage of the top-50 scoring images whose class matches that of the text query, averaged over the 50 test classes.
In \cite{Reed_2016_VisualDescriptions} the best retrieval performance for both datasets (when considering only feature learning) is achieved by having a CRNN directly processing the textual descriptions.
What is also interesting is the substantial performance gain with respect to unsupervised word2vec features.

For the Birds dataset, as an alternative to the textual descriptions, there are manually created fine-grained attributes available for each of the images.
When relying on these attributes Reed et al. report state of the art results on the dataset \cite{Reed_2016_VisualDescriptions} not reached by their text processing neural networks.

In the bottom part of Table \ref{tab:result-zeroshot}, we report the performance of the same architectures optimized using our proposed CCA layer in combination with a pairwise ranking loss.
We observe that the CCA layer is able to improve the performance of both models on both datasets.
The gain in retrieval performance within a model class is largest for the convolution only (CNN) text processing models ($\approx 9$ percentage points for the Flowers dataset and $\approx 6$ for CUB).
For the birds dataset the \emph{Word CNN + CCAL} even outperforms the models relying on manually encoded attributes by achieving an AP@50 of $52.2$.
\begin{table}[ht!]
  \caption{Zero-shot retrieval results on Cub and Flowers.}
  \label{tab:result-zeroshot}
  \centering
  \begin{tabular}{lcc}
    \toprule
    Method 					& Flowers 	& Birds \\
    \midrule
    Attributes \cite{Reed_2016_VisualDescriptions}			& -			& 50.0 \\
    Word2Vec	\cite{Reed_2016_VisualDescriptions}				& 52.1		& 33.5 \\
    Word CNN	\cite{Reed_2016_VisualDescriptions} 				& 56.3		& 43.3 \\
    Word CNN-RNN	\cite{Reed_2016_VisualDescriptions}			& 59.6		& 48.7 \\
    \midrule
    Word CNN + CCAL			& 62.2		& 52.2 \\
    Word CNN-RNN + CCAL  	& 64.0		& 49.8 \\
    \bottomrule
  \end{tabular}
\end{table}

\section{Discussion and Conclusion}
\label{sec:conclusion}
We have shown how to use the optimal projection matrices of CCA
as the weights of an embedding layer within a multi-view neural network.
With this CCA layer, it becomes possible to optimize for a
specialized loss function (e.g., related to a retrieval task)
on top of this, exploiting the correlation properties of a latent space provided by CCA.
As this requires to establish gradient flow through CCA, we formulate it to allow easy computation of the partial derivatives $\frac{\partial \mAstar}{\partial \mathbf{x},\mathbf{y}}$ and $\frac{\partial \mBstar}{\partial \mathbf{x},\mathbf{y}}$
of CCA's projection matrices $\mAstar$ and $\mBstar$ with respect to the input data $\vx$ and $\vy$.
With this formulation, we can incorporate CCA as a building block within multi-modality neural networks
that produces maximally-correlated projections of its inputs.
In our experiments, we use this building block within a cross-modality retrieval setting,
optimizing a network to minimize a cosine distance based pairwise ranking loss of the componentwise-correlated CCA projections.
Experimental results show that when using the cosine distance for retrieval (as is common for correlated views),
this is superior to optimizing a network for maximally-correlated projections (as done in DCCA), or not using CCA at all.
This observation holds in our experiments on a variety of different modality pairs as well as two different retrieval scenarios.

When investigating the experimental results in more detail,
we find that the correlation-based methods (DCCA, CCAL)
consistently outperform the models that learn the embedding projections from scratch.
A direct comparison of DCCA with the proposed CCAL-$\mathcal{L}_{rank}$ reveals two learning scenarios
where CCAL-$\mathcal{L}_{rank}$ is superior:
(1) the low data regime, where we found that the CCA layer acts as a strong regularizer to prevent over-fitting;
(2) when learning the entire retrieval representation (network parameterization) from scratch,
not relying on pre-trained or hand-crafted features (see Section \ref{subsec:audio_sheet}).
Our intuition on this is that incorporating the task-specific retrieval objective already during training 
encourages the networks to learn embedding representations that are beneficial for retrieval at test-time.
This is the important conceptual difference compared to the Trace Norm Objective (TNO) of DCCA,
which does not focus on the retrieval task.
However, when using the CCA layer we also inherit one drawback of the pairwise ranking loss,
which is the additional hyper-parameter (margin $\alpha$) that needs to be determined on the validation set.

Finally, we would like to emphasize that our CCA layer is a general network component which
could provide a useful basis for further research, e.g., as an intermediate processing
step for learning binary cross-modality retrieval representations.

\section*{Acknowledgements}
This work is supported by the Austrian Ministries BMVIT and BMWFW,
and the Province of Upper Austria via the COMET Center SCCH,
and by the European Research Council (ERC) under the EU’s Horizon
2020 Framework Programme (ERC Grant Agreement number 670035,
project "Con Espressione").
The Tesla K40 used for this research was donated by the NVIDIA
corporation.

\bibliography{bibliography}
\bibliographystyle{plain}

\clearpage
\section*{Appendix}
%
\subsection*{Implementation Details}

Backpropagating the errors through the CCA projection matrices is not trivial. The optimal CCA projection matrices are given by $\mAstar = \Sxx^{-1/2} \mU$ and $\mBstar = \Syy^{-1/2} \mV$, where $\mU$ and $\mV$ are derived from the singular value decomposition of $\mT = \Sxx^{-1/2} \Sxy \Syy^{-1/2} = \mU \diag(\vd) \mV'$ (see Section \ref{sec:classic_cca}). The proposed model needs to backpropagate the errors through the CCA transformations, i.e., it requires the gradients of the projected data $\vx^{\!*}={\mAstar}' \vx$ and $\vy^{\!*}={\mBstar}' \vy$ wrt.\ $\vx$ and $\vy$. Applying the chain rule, this further requires the gradients of $\mU$ and $\mV$ wrt.\ $\mT$, and the gradients of $\mT$, $\Sxx^{-1/2}$, $\Sxy$ and $\Syy^{-1/2}$ wrt. $\vx$ and $\vy$. 

The main technical challenge is that common auto-differentiation tools such as Theano \cite{Theano} or Tensor Flow \cite{abadi2016tensorflow} do not provide derivatives for the inverse squared root and singular value decomposition of a matrix.\footnote{Note that this is not relevant for the DCCA model introduced in \cite{Andrew2013DCCA} because it only derives the CCA projections after optimizing the TNO.} To overcome this, we replace the inverse squared root of a matrix by using its Cholesky decomposition as described in \cite{hardoon2004canonical}. Furthermore, we note that the singular value decomposition is required to obtain the matrices $\mU$ and $\mV$, but in fact those matrices can alternatively be obtained by solving the eigendecomposition of $\mT\mT' = \mU \diag(\ve) \mU'$ and $\mT'\mT = \mV \diag(\ve) \mV'$ \cite[Eq.~270]{Cookbook}. This yields the same left and right eigenvectors we would obtain from the SVD (except for possibly flipped signs, which are easy to fix), along with the squared singular values ($e_i = d_i^2$).
Note that $\mT\mT'$ and $\mT'\mT$ are symmetric, and that the gradients of eigenvectors of symmetric real eigensystems have a simple form \cite[Eq.~7]{Magnus1985_eighgrad}. Furthermore, $\mT\mT'$ and $\mT'\mT$ are differentiable wrt.\ $\vx$ and $\vy$, enabling a sufficiently efficient implementation in a graph-based, auto-differentiating math compiler\footnote{The code of our implementation of the CCA layer is available at \repo}.

The following section provides a detailed description of the implementation of the CCA layer.

\subsection*{Forward Pass of CCA Projection Layer}
  For easier reproducibility, we provide a detailed description of the \emph{forward pass} of the proposed CCA layer in Algorithm \ref{nips2017:algo:cca_layer}.
To train the model, we need to propagate the gradient through the CCA layer (backward pass).
We rely on auto-differentiation tools (in particular, \emph{Theano}) implementing the gradient for each individual computation step in the forward pass, and connecting them using the chain rule.


The layer itself takes the latent feature representations (a batch of $m$ paired vectors $\mX \in \mathbb{R}^{d_x \times m}$ and $\mY \in \mathbb{R}^{d_y \times m}$) of the two network pathways $f$ and $g$ as input and projects them with CCA's analytical projection matrices.
At train time, the layer uses the optimal projections computed from the current batch.
When applying the layer at test time it uses the statistics and projections remembered from last training batch (which can of course be recomputed on a larger training batch to get more stable estimate).

\begin{algorithm*}
\caption{Forward Pass of CCA Projection Layer.}
\label{nips2017:algo:cca_layer}
\begin{algorithmic}[1]
  
  \vspace{0.2cm}
  \State Input of layer: $\mX \in \mathbb{R}^{d_x \times m}$ and $\mY \in \mathbb{R}^{d_y \times m}$  			\Comment{hidden representation of current batch}
  \State Returns: $\mXstar$ and $\mYstar$  	\Comment{CCA projected hidden representation}
  \State Parameters of layer: $\mu_{x}$, $\mu_{y}$ and $\mAstar$, $\mBstar$ 	\Comment{means and CCA projection matrices}
  
  \vspace{0.3cm}
  \If{train\_time}  \Comment{\textbf{update statistics and CCA projections during training}}
  	
    \vspace{0.2cm}
    \State $\mu_{x} \leftarrow \frac{1}{m} \sum_{i} \mX_i$	\Comment{update $\mu_{x}$ and $\mu_{y}$ with means of batch}
    \State $\mu_{y} \leftarrow \frac{1}{m} \sum_{i} \mY_i$
    
    \vspace{0.2cm}
    \State $\overbar\mX = \mX - \mu_{x}$	\Comment{mean center data}
    \State $\overbar\mY = \mY - \mu_{y}$
    
    \vspace{0.2cm}
    \State $\Shxx = \frac{1}{m-1} \overbar\mX' \overbar\mX + r \mathbf{I}$	\Comment{estimate covariances of batch}
    \State $\Shyy = \frac{1}{m-1} \overbar\mY' \overbar\mY + r \mathbf{I}$
    \State $\Shxy = \frac{1}{m-1} \overbar\mX' \overbar\mY$
    
    \vspace{0.2cm}
    \State $\mathbf{C}_{xx}^{-1} = \cholesky(\Shxx)^{-1}$  \Comment{compute inverses of Cholesky factorizations} \label{line_cholesky_0}
    \State $\mathbf{C}_{yy}^{-1} = \cholesky(\Shyy)^{-1}$ \label{line_cholesky_1}
    
    \vspace{0.2cm}
    \State $\mT = \mathbf{C}_{xx}^{-1} \Shxy (\mathbf{C}_{yy}^{-1})'$  \Comment{compute matrix $\mT$} \label{line_T}
    
    \vspace{0.2cm}
    \State $\mathbf{e}, \mathbf{U} = \eigen(\mT \mT')$  \Comment{compute eigenvectors of $\mT \mT'$ and $\mT' \mT$} \label{line_eigen_0}
    \State $\mathbf{e}, \mathbf{V} = \eigen(\mT' \mT)$ \label{line_eigen_1}
    
    \vspace{0.2cm}
    \State $\mAstar \leftarrow \mathbf{C}_{xx}^{-1} \mathbf{U}$  \Comment{compute and update CCA projection matrices} \label{line_store_proj}
    \State $\mBstar \leftarrow \mathbf{C}_{yy}^{-1} \mathbf{V}$
  	
    \vspace{0.2cm}
    \State $\mAstar \leftarrow \mAstar \cdot \sgn(\diag({\mAstar}' \Shxy \mBstar))$  \Comment{flip signs of projection matrices} \label{line_flip}
    
  \vspace{0.2cm}
  \Else  \Comment{\textbf{at test time use statistics estimated during training}}
  	
    \vspace{0.2cm}
    \State $\overbar\mX = \mX - \mu_{x}$	\Comment{mean center test data}
    \State $\overbar\mY = \mY - \mu_{y}$
  
  \vspace{0.2cm}
  \EndIf
  
  \vspace{0.2cm}
  \State $\mXstar = \overbar\mX \mAstar$	\Comment{project latent representations with CCA projections} \label{line_project_0}
  \State $\mYstar = \overbar\mY \mBstar$ \label{line_project_1}
  
  \vspace{0.2cm}
  \Return $\mXstar \, \mYstar$
  
\end{algorithmic}
\end{algorithm*}

As not all of the computation steps are obvious, we provide further details for the crucial ones.
In line \ref{line_cholesky_0} and \ref{line_cholesky_1}, we compute the Cholesky factorization instead of the matrix square root, as the latter has no gradients implemented in \emph{Theano}. As a consequence, we need to transpose
$\mathbf{C}_{yy}^{-1}$ when computing $\mT$ in line \ref{line_T} \cite{hardoon2004canonical}.
In line \ref{line_eigen_0} and \ref{line_eigen_1}, we compute two eigen decompositions
instead of one singular value decomposition (which also has no gradients implemented in \emph{Theano}).
In line \ref{line_flip}, we flip the signs of first projection matrix to match the second to only have positive correlations. This property is required for retrieval with cosine distance.
Finally, in line \ref{line_project_0} and \ref{line_project_1}, the two views get projected using $\mAstar$ and $\mBstar$. At test time we apply the projections computed and stored during training (line \ref{line_store_proj}).

\subsection*{Investigations on Correlation Structure}
\begin{figure}[t!]
 \centerline{\includegraphics[width=0.95\columnwidth]{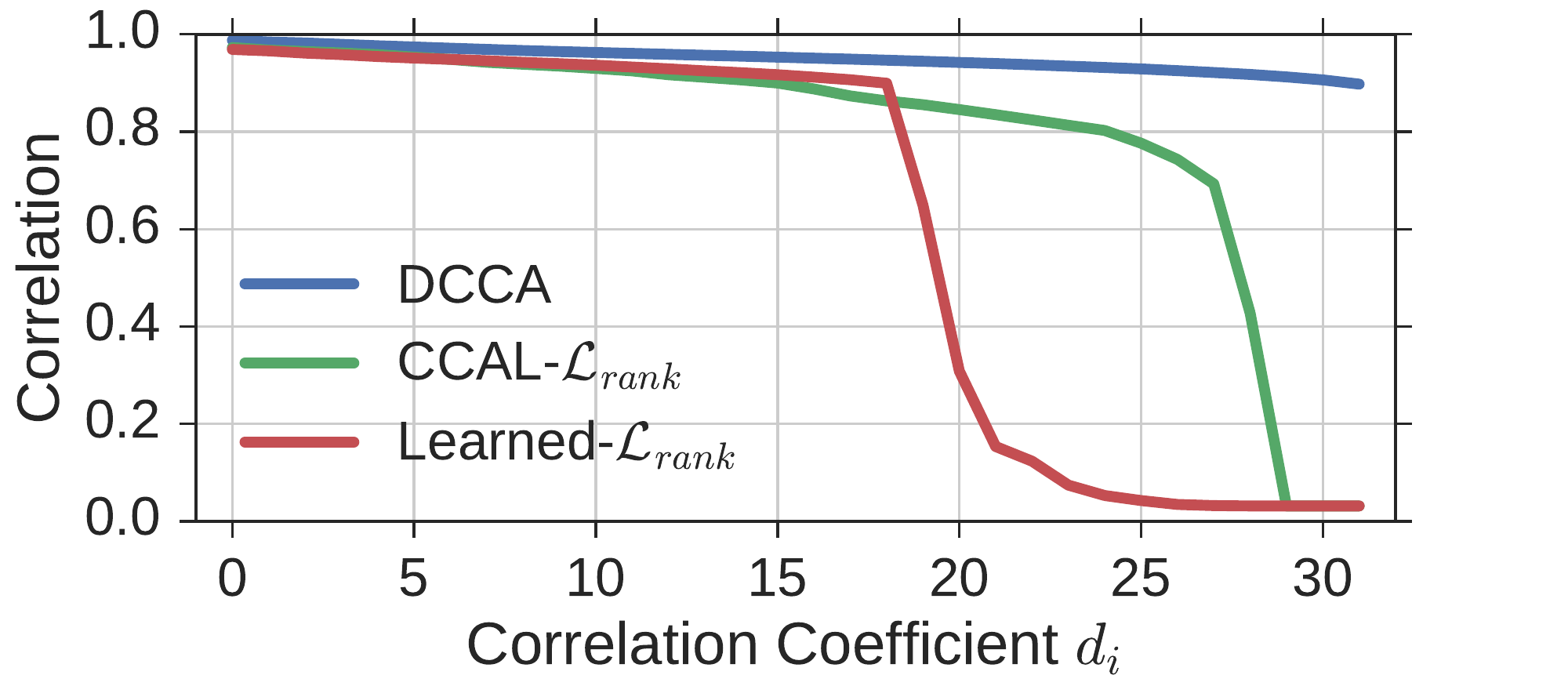}}
 \caption{Comparison of the $32$ correlation coefficients $d_i$ (the dimensionality of the retrieval space is $32$)
          of the topmost hidden representations $\vx$ and $\vy$ of the audio-to-sheet-music dataset and
          the respective optimization paradigm. The maximum correlation possible is $1.0$ for each coefficient}
 \label{fig:comparison_of_correlation}
\end{figure}

As an additional experiment we investigate the correlation structure of the learned representations for all three paradigms.
For that purpose we compute the topmost hidden representation $\vx$ and $\vy$ of the audio-sheet-music-pairs
and estimate the canonical correlation coefficients $d_i$ of the respective embedding spaces.
For the present example this yields $32$ coefficients which is the dimensionality $k$ of our retrieval embedding space.
Figure \ref{fig:comparison_of_correlation} compares the correlation coefficients where $1.0$ is the maximum value reachable.
The most prominent observation in Figure \ref{fig:comparison_of_correlation}
is the high correlation coefficients of the representation learned with \emph{DCCA}.
This structure is expected as the TNO focuses solely on correlation maximization.
However, when recalling the results of Table \ref{tab:result-audio2sheet} we see that this
does not necessarily lead to the best retrieval performance.
The freely learned embedding \emph{Learned-$\mathcal{L}_{rank}$}
shows overall the lowest correlation but achieves comparable results to DCCA on this dataset.
In terms of overall correlation, CCAL-$\mathcal{L}_{rank}$ is situated in-between the two other approaches.
We have seen in all our experiments that combining both concepts in a unified retrieval paradigm
yields best retrieval performance over different application domains as well as data regimes.
We take this as evidence that componentwise-correlated projections support cosine distance based embedding space learning.

\subsection*{Architecture and Optimization}
In the following we proved additional details for our experiments carried out in Section \ref{sec:experiments}.
\subsubsection*{Image-Text Retrieval}
We start training with an initial learning rate of either
$0.001$ (all models on IAPR TC-12 and Flickr30k Learned-$\mathcal{L}_{rank}$) or $0.002$ (Flickr30k DCCA and CCAL-$\mathcal{L}_{rank}$)
\footnote{The initial learning rate and parameter $\alpha$ are determined by grid search on the evaluation measure MRR on the validation set.}.
In addition, we apply $0.0001$ $L2$ weight decay
and set the batch size to $1000$ for all models.
The parameter $\alpha$ of the ranking loss in Equation (\ref{eq:contrastive}) is set to $0.5$.
After no improvement on the validation set for $50$ epochs
we divide the learning rate by $10$ and reduce the patience to $10$.
This learning rate reduction is repeated three times.

\subsubsection*{Audio-Sheet-Music Retrieval}
Table \ref{tab:model_architecture} provides details on our audio-sheet-music retrieval architecture.
\begin{table}[ht]
\caption{\small Architecture of audio-sheet-music model.
BN: Batch Normalization, ELU: Exponential Linear Unit,
MP: Max Pooling, Conv($3$, pad-1)-$16$: $3 \times 3$ convolution, 16 feature
maps and padding 1.}
\vspace*{2mm}
\centering
\footnotesize
\begin{tabular}{c|c}
\hline
Sheet-Image $40 \times 100$ & Spectrogram $136 \times 100$ \\
\hline
$2\times$Conv($3$, pad-1)-$16$	&	$2\times$ Conv($3$, pad-1)-$16$ \\
BN-ELU + MP($2$)					&	BN-ELU + MP($2$) \\
$2\times$Conv($3$, pad-1)-$32$	&	$2\times$ Conv($3$, pad-1)-$32$ \\
BN-ELU + MP($2$)					&	BN-ELU + MP($2$) \\
$2\times$Conv($3$, pad-1)-$64$	&	$2\times$ Conv($3$, pad-1)-$64$ \\
BN-ELU + MP($2$)					&	BN-ELU + MP($2$) \\
$2\times$Conv($3$, pad-1)-$64$	&	$2\times$ Conv($3$, pad-1)-$64$ \\
BN-ELU + MP($2$)					&	BN-ELU + MP($2$) \\
Conv($1$, pad-0)-$32$-BN	& Conv($1$, pad-0)-$32$-BN \\
GlobalAveragePooling 					& GlobalAveragePooling \\
\hline
\multicolumn{2}{c}{Respective Optimization Target} \\
\end{tabular}
\label{tab:model_architecture}
\end{table}

As in the experiments on images and text we optimize our networks using \emph{adam}
with an initial learning rate of $0.001$ and batch size $1000$.
The refinement strategy is the same but no weight decay is applied
and the margin parameter $\alpha$ of the ranking loss is set to $0.7$.

\subsubsection*{Zero-Shot Retrieval}
Table \ref{tab:model_text_cnn} and \ref{tab:model_text_crnn} provide details on the architectures used for our zero-shot retrieval experiments carried out in Section \ref{subsec:zero-shot}.
The general architectures follow Reed et al. \cite{Reed_2016_VisualDescriptions} but are optimized with a pairwise ranking loss in combination with our proposed CCA layer.
The dimensionality of the retrieval space is fixed to $64$
and both models are again optimized with \emph{adam} and a batch size of $1000$.
The learning rate is set to $0.0007$ for the CNN and $0.01$ for the CRNN
and. The margin parameter $\alpha$ of the ranking loss is set to $0.2$.
In addition we apply a weight decay of $0.0001$ on all trainable parameters of the network for regularization.
\begin{table}[ht]
\caption{\small Architecture of Zero-Shot Retrieval CNN.
VS: Vocabulary Size, BN: Batch Normalization, ELU: Exponential Linear Unit,
MP: Max Pooling, Conv($3$, pad-1)-$16$: $3 \times 3$ convolution, 16 feature
maps and padding 1.}
\vspace*{2mm}
\centering
\footnotesize
\begin{tabular}{c|c}
\hline
ImagenNet Feature $1024$ & Text $VS \times 30 \times 1$ \\
\hline
FC(1024)-BN-ELU	&	$1\times$Conv($3$, pad-\emph{same})-$256$ \\
FC(1024)-BN-ELU	&	BN-ELU + MP($3, 1$) \\
FC(64)			&	$2\times$Conv($3$, pad-\emph{valid})-$256$ \\
				&	FC(1024)-BN-ELU \\
				&	FC(64) \\
\hline
\multicolumn{2}{c}{Respective Optimization Target} \\
\end{tabular}
\label{tab:model_text_cnn}
\end{table}
\begin{table}[ht]
\caption{\small Architecture of Zero-Shot Retrieval CRNN.
VS: Vocabulary Size, BN: Batch Normalization, ELU: Exponential Linear Unit,
MP: Max Pooling, Conv($3$, pad-1)-$16$: $3 \times 3$ convolution, 16 feature
maps and padding 1. GRU-RNN: Gated Recurrent Unit \cite{Chung2014GRU}}
\vspace*{2mm}
\centering
\footnotesize
\begin{tabular}{c|c}
\hline
ImagenNet Feature $1024$ & Text $VS \times 30 \times 1$ \\
\hline
FC(1024)-BN-ELU	&	$1\times$Conv($3$, pad-\emph{same})-$256$ \\
FC(1024)-BN-ELU	&	BN-ELU + MP($3, 1$) \\
FC(64)			&	$2\times$Conv($3$, pad-\emph{valid})-$256$ \\
				&	GRU-RNN(512) \\
				& 	TemporalAveragePooling \\
				&	FC(64) \\
\hline
\multicolumn{2}{c}{Respective Optimization Target} \\
\end{tabular}
\label{tab:model_text_crnn}
\end{table}

\end{document}